# Connection between ambient-pressure and pressure-induced superconducting phases in alkaline iron selenide superconductors

Dachun Gu[1], Liling Sun[1,2]†, Qi Wu[1], Peiwen Gao[1], Jing Guo[1], Chao Zhang[1], Shan Zhang[1], Yazhou Zhou[1], and Zhongxian Zhao[1,2]†

[1]Institute of Physics and Beijing National Laboratory for Condensed Matter Physics, Chinese Academy of Sciences, Beijing 100190, China
[2]Collaborative Innovation Center of Quantum Matter, Beijing, 100190, China

A unique platform for investigating the correlation between the antiferromagnetic (AFM) and superconducting (SC) states in high temperature superconductors is established by the discovery of alkaline iron selenide superconductors which are composed of two spatially separated phases, an AFM insulating phase and a SC phase. Our previous studies show that pressure can fully suppress the superconductivity of ambient-pressure superconducting phase (SC-I) and the AFM long-ranged order concomitantly, then induce another superconducting phase (SC-II) at higher pressure. Consequently, the connection between these two superconducting phases becomes an intriguing issue. In this study, we find a pressure-induced reemergence of superconductivity in $Rb_{0.8}Fe_{2-y}Se_{2-x}Te_x$ ($x$=0, 0.19 and 0.28) superconductors, and reveal that the superconductivity of the SC-I and SC-II phases as well as the AFM long-ranged order state can be synchronously tuned by Te doping and disappear all together at the doping level of $x$=0.4. These results lead us to propose that the two superconducting phases are connected by the AFM phase, *i.e.*, the state of long-ranged AFM order plays a role in giving rise to superconductivity of the SC-I phase, while the state of fully suppressed AFM long-ranged order by pressure drives the emergence of SC-II phase. The versatile roles of AFM states in stabilizing and developing superconductivity in the alkaline iron selenide superconductors are comprehensively demonstrated in this study.

The discovery of superconductivity in $A_{0.8}Fe_{2-y}Se_2$ compounds (A=K, TlRb) [1,2] with $\sqrt{5}\times\sqrt{5}$ arrangement of ordered Fe vacancies (defined as A-245 superconductors thereafter) opened a new exciting avenue for investigations of correlations among superconductivity, antiferromagnetic (AFM) order and lattice structure in Fe-based superconductors [3-9]. Soon after, the superconductivity was also found in Rb-245 and Cs-245 compounds [10-15]. Then, the characteristics of lattice, electronic and magnetic structures have been reported widely for these A-245 superconductors, such as the existence of phase separation [16-19], the ordered Fe vacancy with $\sqrt{5}\times\sqrt{5}$ arrangement in the lattice and its correlation to the AFM order [20-24], absence of hole pockets at the Fermi surface [25-27], the temperature-induced orbital selection [28, 29], all of which are not shown either in the iron arsenide or copper oxide superconductors, therefore the complexity of understanding its superconducting mechanism is raised to a new level. Previous studies found that applying external pressure on A-245 superconductors can suppress the superconductivity of the ambient-pressure superconducting phase (SC-I) [30-32] and induce an emergence of superconductivity in a new superconducting phase (SC-II) [33]. Experimental evidences through comprehensive measurements have exhibited that the pressure-induced SC-II phase is probably driven by a quantum critical phase transition where the host sample undergoes a conversion from a AFM state to a paramagnetic (PM) state [30,34], suggesting that the superconductivity of the SC-II phase may closely tie with the AFM fluctuation [35, 36]. The results reported by high pressure studies attract much attention [37-40], meanwhile, a puzzle about whether the SC-I and SC-II phases are intrinsically connected each other is raised. The answer for it may be helpful to understanding the superconducting mechanism in this unique family of Fe-based superconductors.

It is known that the iso-valence substitution of Te or S, with larger or smaller ionic radius, for Se can distort the lattice, thus impact the state of long-ranged AFM order and corresponding superconductivity in the SC-I phase [41-44]. However, such a doping effect cannot develop the SC-II phase of the A-245 superconductors, even doped to the maximum of the solid solubility. Consequently, there is a growing need to find a more effective fashion that can tune the evolution of superconductivity in both of the SC-I and SC-II phases, and then reveal a possible link between superconductivity in each superconducting phase and the state of AFM order. In this study, we combine two tuning ways, doping Te on Se sites and applying external pressure, to perform a comprehensive investigation on Rb-245 superconductors.

Single crystals of Rb-245 superconductors were grown out by the self-flux method, as reported in Ref. [41]. The actual chemical compositions of all samples investigated were $Rb_{0.8}Fe_{1.7}Se_2$, $Rb_{0.8}Fe_{1.65}Se_{1.8}Te_{0.19}$, $Rb_{0.8}Fe_{1.63}Se_{1.72}Te_{0.28}$ and $Rb_{0.8}Fe_{1.66}Se_{1.6}Te_{0.4}$, respectively, which are identified by the inductive coupled plasma-atomic emission spectrometer.

The *in-situ* high-pressure electrical resistance and *ac* susceptibility measurements were carried out in a home-built refrigerator which was inserted into a nonmagnetic diamond anvil cell. Diamond anvils of 500 μm and 300 μm flats were used for this study. To achieve qausi-hydrostatic pressure environment for the sample, NaCl powders were employed as pressure medium for the high-pressure resistance measurements. High-pressure *ac* susceptibility measurements were conducted using home-made coils that were wound around a diamond anvil [33, 45]. The nonmagnetic rhenium gasket with 200 μm and 100 μm diameter sample holes was used for different runs of high-pressure resistance and magnetic measurements. Pressure was determined by ruby fluorescence method [46]. Temperature was measured with a calibrated Si-diode attached to the diamond anvil cell with accuracy less than 0.1 K.

Figure 1a shows the resistance (R) of an un-doped Rb-245 superconducting sample as a function of temperature (T) at different pressures. The data remarkably demonstrate that a hump in the normal-state resistance, which has been identified to be related to the AFM long-ranged order [24,30], is suppressed significantly upon increasing pressure, the same as that seen in pressurized K-245 and Tl(Rb)-245 superconductors [30, 32]. At pressure ~8.4 GPa, we found that the resistance hump becomes almost featureless, which indicates that the long-ranged AFM order is destructed at this pressure according to the experimental results observed by high pressure X-ray [30] and neutron diffraction measurements [34]. Zooming in the plot of R-T curve in the low temperature range, the pressure-induced decrease in $T_c$ is found (Fig.1b). At 7.2 GPa, the superconductivity is fully suppressed, and then a pressure-induced new drop appears in the pressure range from 8.4 to 11.8 GPa (Fig.1c). With further increasing pressure to 14.1 GPa, the resistance drop vanishes, similar to that seen in other A-245 superconductors [33]. To more fully characterize the superconducting state in the pressurized Rb-245 superconductor, we performed *ac* susceptibility measurements at pressures of 1.1 GPa, 3.5 GPa and 11 GPa, which fall in the pressure regime of SC-I and SC-II phases, respectively. The results show that the host sample are diamagnetic, so compelling that the sample is superconducting at these three pressure points (Fig.1d and 1e ). Further measurements under magnetic field and dc current for the sample observe a shift of the R-T curve to lower temperature, demonstrating that pressure indeed induces a presence of the SC-II phase (Fig.1f and 1g). This is the first observation of pressure-induced reemergence of superconductivity in the Rb-245 superconductor.

Next we performed high-pressure studies on the Te-doped Rb-245 superconductors. We found that the resistance hump also exists in the pressure-free $Rb_{0.8}Fe_{2-y}Se_{2-x}Te_x$ ($x$=0.19 and 0.28) (Fig.2a and 2f). Applying external pressure yields a dramatic suppression of the resistance hump in both of the samples. Careful inspection on their R-T plots in the lower temperature range, it is seen that the $Tc$ of the SC-I phase is declined with increasing pressure (Fig.2b and Fig.2g). Upon further increasing pressure, the drops featuring the SC-II phase show up at 11.5 GPa for the $x$=0.19 sample and at 12.4 GPa for the $x$=0.28 sample, respectively (Fig.2c and Fig.2h). The superconducting transitions of the SC-II phases are confirmed by shifts of the R-T curve to the lower temperature when the magnetic field or current is increased (Fig 2d, 2e, 2i and 2j).

Figure 3 represents temperature dependence of resistance for the $x$=0.4 sample at different pressures. Notably, heavier doping results in a semiconducting behavior, though its long-ranged AFM order state is fully suppressed and the sample is in a paramagnetic state [41]. With increasing pressure, the semiconducting behavior is suppressed dramatically (Fig.3b-3c). At 13 GPa and above, its resistance decreases remarkably with lowing temperature (Fig.3d), indicating that the sample moves into a metallic state. No SC-II phase is detected in the sample under pressure up to 15.5 GPa (Fig.3e).

The overall behavior of Rb-245 superconductors is summarized in the electronic phase diagram of pressure-composition-temperature, as shown in Fig.4. Adopting pressure as a control parameter, the $Tc$ of the SC-I phase in the $x=0$ sample goes down with increasing pressure, and a new superconducting phase (SC-II) emerges within 8.4 GPa-11.8 GPa, after the SC-I phase is fully suppressed. The maximum onset $Tc$ of the SC-II phase is ~53 K at 11.8 GPa. The diagram with double superconducting phases have been observed in K-245 and Tl(Rb)-245 superconductors [33], so that the results reported in this study further indicate that the pressure-induced reemergence of superconductivity is a common phenomenon for the family of alkaline iron selenide superconductors.

Regarding $x=0.19$ and 0.28 samples, the ambient-pressure value of the $Tc$ in their SC-I phase ($Tc=29.8$ K for the $x=0.19$ sample and 24.2 K for the $x=0.28$ sample) is lower than that ($Tc=33$ K) of the un-doped sample, implying that the Te-doping is not in favor of the superconductivity [41-43]. Remarkably, the reduced $Tc$ of the SC-I phase in the $x=0.19$ and 0.28 samples can be partially recovered by 1.1 K for the $x=0.19$ sample at 1.2 GPa, and by 2.2 K for the $x=0.28$ sample at 1.7 GPa, respectively. Our results suggest that the $Tc$ of the SC-I phase in the A-245 superconductors is very sensitive to lattice distortion, and that duel tuning fashions of chemical doping and applying external pressure are effective in uncovering the relationship between superconductivity and AFM order phase.

It is worth noting that neither the SC-I nor SC-II phases is seen in the $x$=0.4 sample, at the doping level of which the long-ranged AFM order is fully suppressed at ambient pressure [30, 34, 36, 41] (inset of Fig.4). Interestingly, when the paramagnetic semiconducting sample ($x$=0.4) is pressurized into a metallic state, the SC-II phase is not observed (Fig.3d-3e and Fig.4). These results suggest that the long-ranged AFM order is essential to maintain the existence of the SC-I phase, in good agreement with results obtained from high-pressure neutron studies on the TlRb-245 superconductor [34], an analogue of the Rb-245 superconductor. Previous high-pressure studies on K-245 and TlR-245 superconductors reveal that the SC-II phase emerges from a metallic state, and is driven by a quantum critical transition where the long-ranged AFM order is fully suppressed [30]. While, the x=0.4 sample presents a paramagnetic semiconducting state at ambient pressure, as a results that it is impossible for pressure to tune such a heavy-doped sample into a superconducting state, even though the sample is compressed into a metallic state.

In conclusion, we find the pressure-induced reemergence of superconductivity in $Rb_{0.8}Fe_{2-y}Se_{2-x}Te_x$ ($x$=0, 0.19 and 0.28) superconductors, and investigate the connection between the SC-I and SC-II phases. We find that Te-doping can significantly suppress the superconductivity in both of the SC-I and SC-II phases and eliminate the two superconducting phases at $x$=0.4. We propose that both of the SC-I and SC-II phases connect to the same AFM phase, *i.e.* the AFM long-ranged order state stabilize the superconductivity of SC-I phase, while the pressure-induced quantum critical transition, resulted from a full suppression of AFM long-ranged order state, drives the reemergence of superconductivity of SC-II phase. These findings may provide a panorama picture on the pressure and doping plane for the superconducting behaviors of the A-245 superconductors and enrich the knowledge for the roles of AFM long-ranged order state in high temperature superconductors.

**Acknowledgements**

We thank Prof. Jianqi Li for valuable discussions. This work has been supported by the NSF of China (Grant No. 91321207 and 11427805), 973 projects (Grant No.2011CBA00100 and 2010CB923000) and the Strategic Priority Research Program (B) of the Chinese Academy of Sciences (Grant No. XDB07020300).

46. H. K. Mao, J. Xu, and P. M. Bell, J. Geophys. Res. **91,** 4673-4676 (1986).

†To whom correspondence should be addressed.

E-mail: llsun@iphy.ac.cn and zhxzhao@iphy.ac.cn .

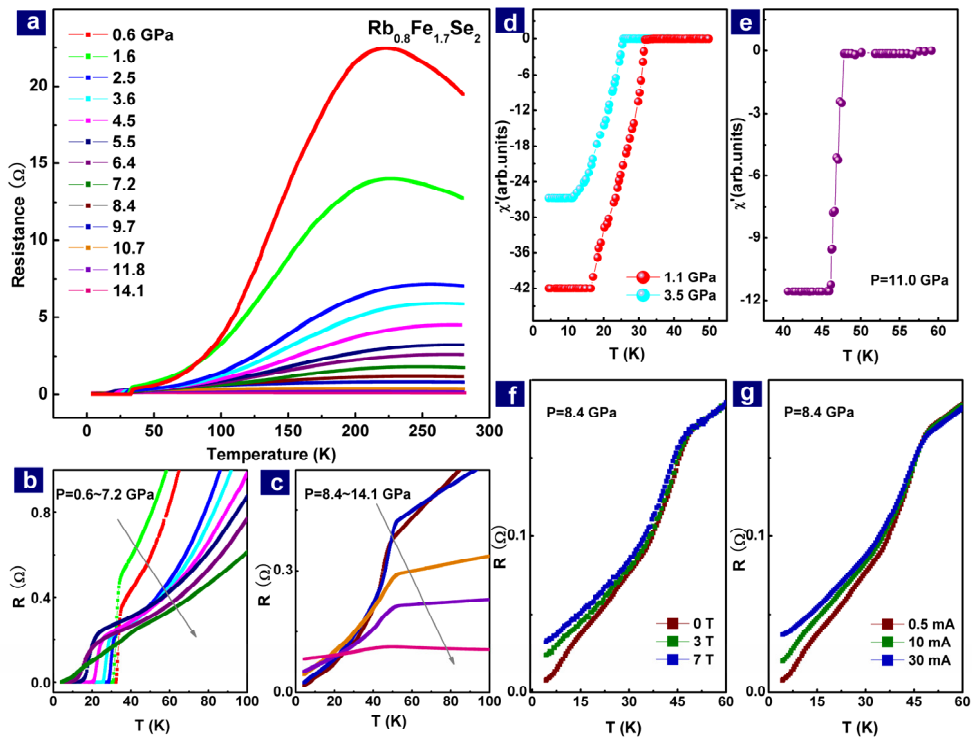

Figure 1 (Color online) Resistance as a function of temperature for $Rb_{0.8}Fe_{1.7}Se_2$ measured under different conditions. (a) Temperature dependence of resistance at different pressures (b) Pressure-induced suppression of the superconducting transition temperature in the pressure range of 0.6-7.2 GPa. (c) Illustration of pressure-induced resistance drop and its change with increasing pressure from 8.4 GPa to 14.1 GPa. (d) and (e) *ac* susceptibility measurements at 1.1 GPa, 3.5 GPa and 11 GPa, which fall in the pressure regime of SC-I and SC-II phases, indicating diamagnetic signals at 31.1 K, 25.6 K and 47.8 K, respectively. (f) and (g) The shift of resistance-temperature curve to lower temperature upon increasing magnetic field and current, indicating that the pressure-induced resistance drop is related to a superconducting transition.

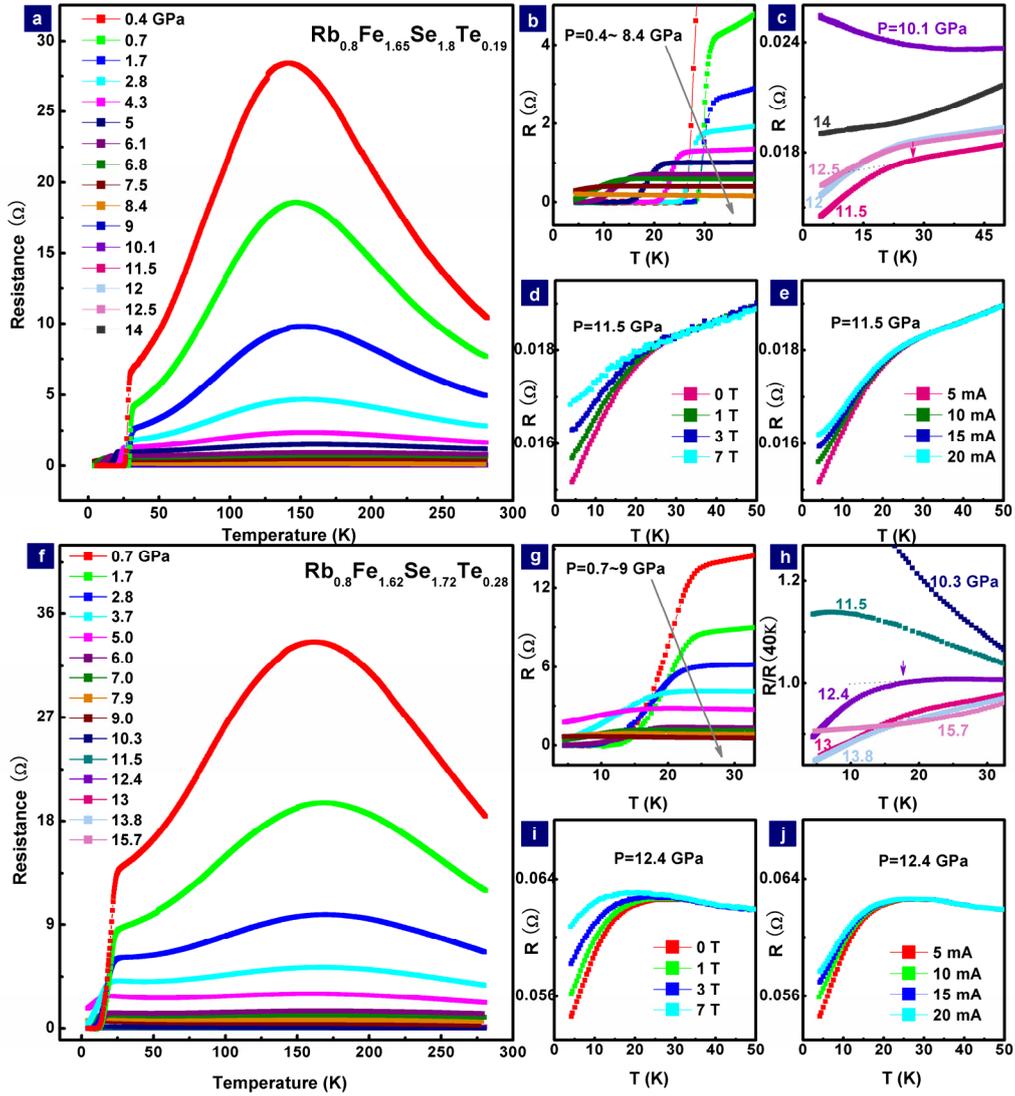

Figure 2 (Color online) Temperature-dependent resistance of Te-doped Rb-245 superconductors at different pressures. (a) Typical resistance-temperature (R-T) curves of single crystal of $Rb_{0.8}Fe_{1.65}Se_{1.8}Te_{0.19}$ at pressures up to 14 GPa. (b) The changes in superconductivity of the SC-I phase with increasing pressure. A full suppression of the superconductivity is found at 8.4 GPa. (c) Pressure-induced reemergence of superconductivity at higher pressure. (d) and (e) The superconducting transition temperature ($T_c$) of the pressure-induced SC-II phase is suppressed by increasing magnetic field and current. (f) Typical R-T curves of single crystal of $Rb_{0.8}Fe_{1.63}Se_{1.72}Te_{0.28}$ at pressures up to 15.7 GPa. (g) The enlarged view of R-T curves of the $x$=0.28 sample at different pressures. (h) Pressure-induced reemergence of superconductivity starting at 11.5 GPa and ending at 15.7GPa. (i) and (j) The $T_c$ of the SC-II phase in the $x$=0.28 is suppressed by increasing magnetic field and current.

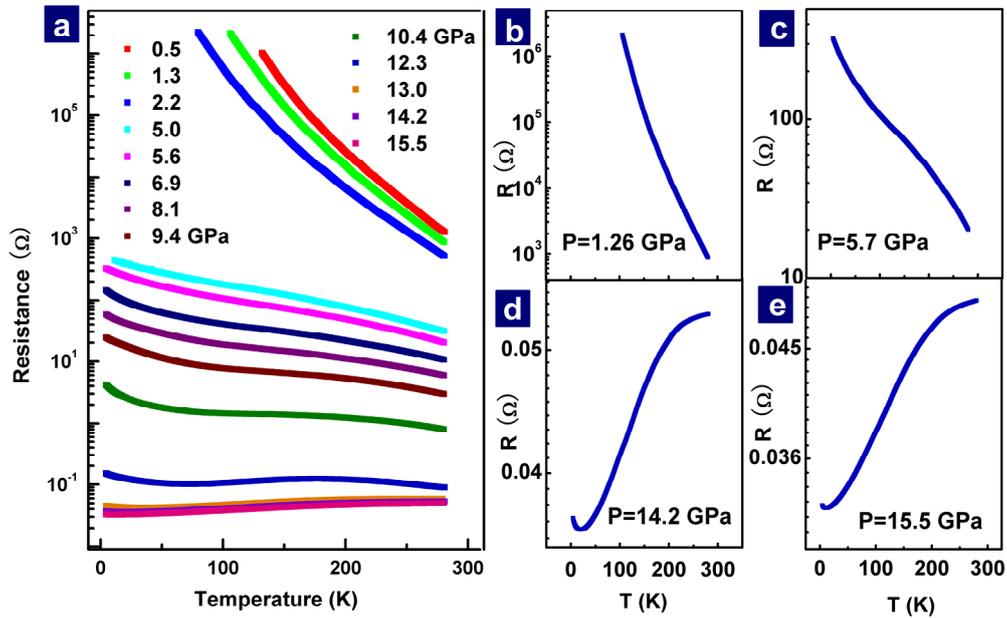

Figure 3 (Color online) Temperature dependence of resistance at different pressures in $Rb_{0.8}Fe_{1.66}Se_{1.6}Te_{0.4}$ sample. (a) R-T curves measured in the pressure range of 0.5 GPa - 15.5 GPa, showing a pressure-induced remarkable suppression of the insulating behavior. (b)-(e) R-T curves measured at 2.2 GPa, 5.7 GPa, 14.2 and 15.5 GPa,

respectively.

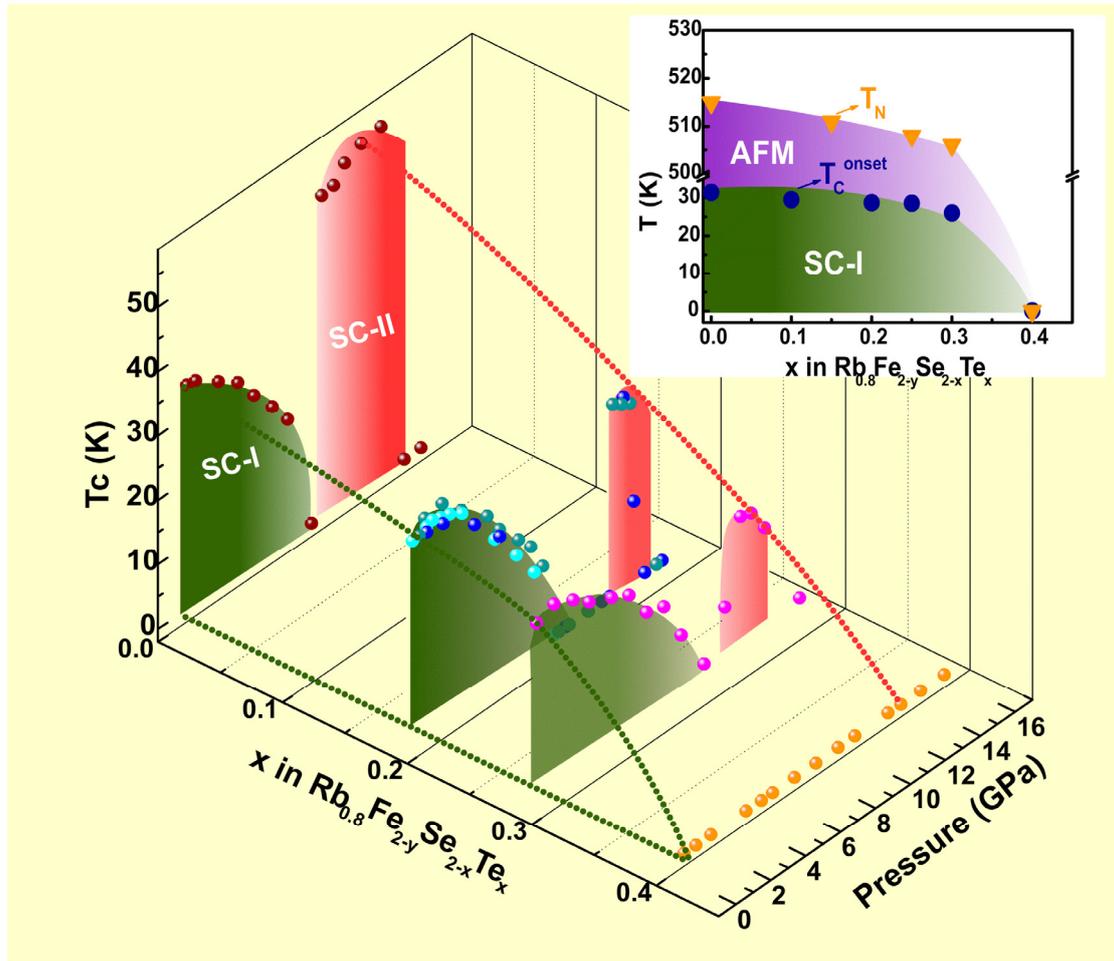

Figure 4 (Color online) The phase diagram of temprature-doping-pressure of Rb-245 samples. Te doping supresses the superconductivity of SC-I and SC-II phases. As the doping level reaches 0.4, SC-I and SC-II disappear together in the pressure range investigated, demonstarting an intimate connection between the two superconducting phases. The inset of the main figure is taken from Ref. [41]. The green and red dotted lines guide to eye.